
%

\input harvmac.tex
\input epsf.tex
\noblackbox
\Title{\vbox{\baselineskip12pt\hbox{CLNS 92/1148, USC-92/010}}}
{\vbox{\centerline{S-matrices for Perturbed N=2 Superconformal Field
Theory }
\vskip2pt\centerline{from  Quantum Groups }}}
\centerline{A. LeClair$^{\dagger}$ }
\bigskip
\centerline{Newman Laboratory}
\centerline{Cornell University}
\centerline{Ithaca, NY 14853}
\bigskip
\centerline{D. Nemeschansky$^*$ and N.P. Warner$^{* \dagger}$
{\abstractfont
\footnote{}{$^*$ Work supported in part by funds provided by the DOE
under grant No. DE-FG03-84ER40168.}
\footnote{}{$^\dagger$ Alfred P. Sloan Foundation Fellow.}}}
\bigskip
\centerline{Physics Department}
\centerline{University of Southern California}
\centerline{University Park}
\centerline{Los Angeles, CA 90089-0484.}
\vskip 1.0cm

S-matrices for  integrable perturbations of
$N=2$ superconformal field theories are studied.
The models we consider correspond to perturbations
of the coset theory $G_k \times H_{g-h} /H_{k+g-h} $.
The perturbed models are closely related to $\hat G$-affine Toda
theories with a background charge tuned to $H$.
Using the quantum group restriction of the affine Toda theories
we derive the S-matrix.

\Date{5/92}
%
%
%
%
%
%
%
%

%
%
%
%

%
%
%

%
%

\def\tilde{\widetilde}
\def\bar{\overline}
\def\hat{\widehat}
\def\*{\star}
\def\[{\left[}
\def\]{\right]}
\def\({\left(}		
\def\){\right)}

%
%
\def\rhh#1{ { \hat{\rho}^{(H)}_{#1} }}
\def\av#1{{\vec{\al}_{#1} } }
\def\k#1{{ K^{(#1)} }}
\def\rh#1{{ \hat{\rho}_{#1} }}
\def\v#1{{ V_{#1} }}
\def\w#1{{ W_{#1} }}

\def\zb{{ \bar{z} }}
\def\qb{\bar{Q}}
\def\Q#1{{ Q_{#1} }}
\def\Qb#1{{ \qb_{#1} }}
\def\lh{{ \lambda^{(H)} }}
\def\lhi#1{{ \lambda^{(H)}_#1 }}
\def\co#1#2{ G_#1 \ot H_#2 / H_{#1 +#2} }
\def\gh{{\hat{G}}}
\def\ot{\otimes}
\def\zb{{\bar{z} }}
\def\frac#1#2{{#1 \over #2}}
\def\inv#1{{1 \over #1}}
\def\half{{1 \over 2}}
\def\d{\partial}

\def\2pi{\hbox{$2\pi i$}}

\def\dsl{\raise.15ex\hbox{/}\kern-.57em\partial}
\def\Dsl{\,\raise.15ex\hbox{/}\mkern-.13.5mu D}
%
%
\def\th{\theta}		
		
\def\be{\beta}
\def\al{\alpha}

\def\la{\lambda}	
\def\de{\delta}

%
%
\def\CA{{\cal A}}		
		\def\CF{{\cal F}}

\def\CM{{\cal M}}		
		
		\def\CU{{\cal U}}

\def\2pi{\hbox{$2\pi i$}}

\def\dsl{\raise.15ex\hbox{/}\kern-.57em\partial}
\def\Dsl{\,\raise.15ex\hbox{/}\mkern-.13.5mu D}
%
%
\def\phib{\bar{\phi}}
%
\font\numbers=cmss12
\font\upright=cmu10 scaled\magstep1
\def\stroke{\vrule height8pt width0.4pt depth-0.1pt}
\def\topfleck{\vrule height8pt width0.5pt depth-5.9pt}
\def\botfleck{\vrule height2pt width0.5pt depth0.1pt}
\def\Zmath{\vcenter{\hbox{\numbers\rlap{\rlap{Z}\kern
0.8pt\topfleck}\kern
2.2pt
                   \rlap Z\kern 6pt\botfleck\kern 1pt}}}
\def\Qmath{\vcenter{\hbox{\upright\rlap{\rlap{Q}\kern
                   3.8pt\stroke}\phantom{Q}}}}
\def\Nmath{\vcenter{\hbox{\upright\rlap{I}\kern 1.7pt N}}}
\def\Cmath{\vcenter{\hbox{\upright\rlap{\rlap{C}\kern
                   3.8pt\stroke}\phantom{C}}}}
\def\Rmath{\vcenter{\hbox{\upright\rlap{I}\kern 1.7pt R}}}
\def\Z{\ifmmode\Zmath\else$\Zmath$\fi}
\def\QQ{\ifmmode\Qmath\else$\Qmath$\fi}
\def\N{\ifmmode\Nmath\else$\Nmath$\fi}
\def\C{\ifmmode\Cmath\else$\Cmath$\fi}
\def\R{\ifmmode\Rmath\else$\Rmath$\fi}


\def\LG{Lan\-dau-Ginz\-burg\ }

\def\mkl{{\cal M}_{k,\ell} (G;H)}
\def\Gkl#1{{{G_k \times #1_\ell} \over {#1_{k+\ell}}}}
\def\UU{{\cal U}} \def\half{{1 \over 2}}

\def\primh{\Phi^{\Lambda, \lambda_+}_{\lambda_-}}
\def\nup#1{{\it Nucl.\ Phys.} \ {\bf B#1\/}}
\def\ijmp#1{{\it Int.\ J. \ Mod. \ Phys.} \ {\bf A#1\/}}
\def\plt#1{{\it Phys.\ Lett.}\ {\bf#1B\/}}
\def\lmp#1{{\it Lett.\ Math. \ Phys. }\ {\bf#1\/}}
\def\smd#1{{\it Sov.\ Math. \ Dokl. }\ {\bf#1\/}}
\def\cmp#1{{\it Commun. \ Math. \ Phys.} \ {\bf #1\/}}
\def\prl#1{{\it Phys.\ Rev. \ Lett.}\ {\bf#1\/}}
\def\mpl#1{{\it Mod. \ Phys.\ Lett.}\ {\bf#1\/}}
  \def\Gminus{G_{-{1 \over 2}}^-}
\def\Gplus{G_{-{1 \over 2}}^+}

\def\inbar{\vrule height1.5ex width.4pt depth0pt}
\def\IC{\relax\,\hbox{$\inbar\kern-.3em{\rm C}$}}
\def\IP{\relax{\rm I\kern-.18em P}}
\def\IR{\relax{\rm I\kern-.18em R}}
\font\sanse=cmss12
\def\ZZ{\relax{\hbox{\sanse Z\kern-.42em Z}}}
\def\ckpR{\check R^\prime}
\def\uqH{U_q(H)}
\def\uqG{U_q(G)}
\def\uqGh{U_q(\hat G)}
\def\uqHp{U_q(H^\prime)}


\def\BLnlc{D. Bernard and A. LeClair, \cmp{142} (1991) 99.}
\def\PasS{V. Pasquier and H. Saleur, \nup{330} (1990) 523.}
\def\ntwo{P. Fendley, S. D. Mathur, C. Vafa, and N. P. Warner,
\plt{243} (1990) 257.}
\def\FLMW{P. Fendley, W. Lerche, S. D. Mathur,
and N. P. Warner, \nup{348} (1991) 66.}
%
\def\Nak{T. Nakatsu,  \nup{356} (1991) 499.}
\def\VeFa{H. J. de Vega and V. A. Fateev,
\ijmp{6} (1991) 3221.}
\def\Jimborev{M. Jimbo, \ijmp{4} (1989) 3759.}
\def\Hollow{T. Hollowood, `Quantizing SL(N) Solitons and the Hecke
Algebra', Oxford preprint OUTP-92-03P, and OUTP-90-15P .}
\def\KaTh{M. Karowski and H. J. Thun, \nup{190} (1981) 61.}
\def\fusion{P. P. Kulish, N. Yu. Reshetikhin, and E. K. Sklyanin,
\lmp{5} (1981) 393. }
\def\fssg{D. Bernard and A. LeClair,  \plt{247} (1990) 309.}
\def\ABL{C. Ahn, D. Bernard, and A. LeClair, \nup{346} (1990) 409. }
\def\cosetFF{D. Kastor, E. Martinec, and Z. Qiu, \plt{200}
(1988) 434\semi
J. Bagger, D. Nemeschansky, and S. Yankielowicz, \prl{60}
(1988) 389\semi
F. Ravanini, \mpl{A3} (1988) 397\semi
P. Christe and F. Ravanini, \ijmp{4} (1989), 897.}
\def\ORW{E. Ogievetsky, N. Yu. Reshetikhin, and P. Wiegmann,
 \nup{280} [FS18] (1987) 45.}
\def\OW{E. Ogievetsky and P. Wiegmann, \plt{168} (1986),  360.}


\def\BLii{D. Bernard and A. LeClair,  \nup{340} (1990) 721.}
\def\rsg{A. LeClair, \plt{230} (1989) 103;  Cornell preprint
CLNS 89/955, in  proceedings of the 1989 Banff workshop on
`Physics, Geometry,  and Topology', Plenum Press,  H.~C.~Lee, editor.
}
\def\RS{N. Yu. Reshetikhin and F. Smirnov,  \cmp{131} (1990) 157. }
\def\Zamoiv{A. B. Zamolodchikov, Sept 1989 preprint, unpublished. }
\def\Jimboii{M. Jimbo, \cmp{102} (1986) 537.}

\def\EY{T. Eguchi and S.-K. Yang,
\plt{224} (1989) 373.}
\def\HM{T. Hollowood and P. Mansfield,
\plt{226} (1989) 73.}
\def\Zamoiii{A. B. Zamolodchikov, \ijmp{4} (1989) 4235;
in {\it Adv. Studies in Pure Math.} vol. {\bf 19 } (1989) 641.}
%

%
%
%
%
%
%

\newsec{{\bf Introduction} }

Over the past few years many new integrable quantum field theories
in two dimensions have been constructed as perturbations of
conformal field theories.   This program was initiated by
A. Zamolodchikov \ref\rzamoi{\Zamoiii}.  Of special interest
are the $N=2$ superconformal theories with perturbations that
preserve the $N=2$ supersymmetry.  For such models the \LG structure
and Bogomolnyi bounds of the superalgebra can be used to deduce much
about the soliton spectrum \ref\rntwo{\ntwo}\ref\rlg{W. Lerche
and N. P. Warner, \nup{358} (1991) 571.}.  Integrable perturbations
of the minimal series of models (with central charge
$c= 3k/(k+2)$ ) were studied in \rntwo, where it
was shown that there are three different integrable perturbations,
corresponding to perturbation by the least relevant, most
relevant, and next-to-most relevant chiral primary fields.
In \rntwo\ some features of the spectrum for the
most relevant perturbations were proposed based on the
\LG formulation of the theories.  More general
$N=2$ supersymmetric integrable models were obtained in
\ref\rflmw{\FLMW} and further results about the
soliton spectrum were obtained using the \LG picture in \rlg.

In \ref\rfssg{\fssg} the S-matrices for the least relevant
perturbations
of the minimal series were proposed based on the quantum group
symmetries
that exist in the models.  Indeed, the $N=2$ supersymmetry was
understood
as a special case of the quantum affine symmetry $U_q(\hat{SU(2)})$,
which occurs at $q=-i$.   Further confirmation
of these S-matrices was provided by a thermodynamic Bethe ansatz
analysis \ref\rfi{P.~Fendley and K.~Intrilligator,`Scattering
and Thermodynamics of Fractionally Charged Supersymmetric Solitons,'
Harvard and Boston preprints (1991) HUTP-91/A043, BUHEP-91-17. }.
Also in \rfi\
the relation between the spectrum proposed in \rfssg\ and the
Landau-Ginzburg formulation of the theories was clarified.
The S-matrices for the most relevant perturbations of the minimal
series of $N=2$ theories and other related models were proposed
in \ref\rfii{P.~ Fendley and K.~Intrilligator,  `Scattering
and Thermodynamics in Integrable $N=2$ Theories,'
Harvard and Boston preprints (1992) HUTP-91/A067, BUHEP-92-5. },
again based largely on
the Landau-Ginzburg picture of the soliton spectrum.  Here it was
found that the solitons generally had fractional fermion number,
and this fact was essential for obtaining the correct S-matrices.
In the quantum group approach, these fractional fermion numbers
are automatically incorporated.

In this paper we use the restricted quantum group approach
\ref\rrsg{\rsg} \ref\rRS{\RS}\  to derive
the S-matrices for a class of integrable perturbations of
$N=2$ superconformal field theories.  We will actually solve a
larger class of models corresponding to perturbations of the
cosets $\co k l $.  The $N=2$ supersymmetric theories
can essentially be obtained from these models by taking
$l = g - h$, where $g$ and $h$ are the dual coxeter numbers of
$G$ and $H$ respectively\ref\rks{Y. Kazama and H. Suzuki,
\nup{234} (1989) 73.}.  The main
features of this approach can be summarized as follows.  Consider
first the situation where $G=H$, which was studied in
\rfssg\ref\rabl{\ABL}\ref\rnlc{\BLnlc}.
For $k=1$, one begins with the $\gh$-affine Toda field theory, with
zero background charge. (The affine extension of $G$ will be
denoted by $\gh$.)
This field theory can be solved by using the affine quantum group
symmetry, $\uqGh$,  that exists in the model \rnlc. The coset model
is obtained by turning on a background charge for the Toda fields.
This background charge modifies the conformal dimension of the
conserved charges that generate the finite quantum subalgebra,
$\uqG$, of $\uqGh$.  As a result the conserved charges
become dimension zero screening
charges.  One then uses the $\uqG$ symmetry algebra of screening
charges to restrict the S-matrices of the $\gh$-affine Toda theory
to obtain the S-matrices of the perturbed coset theory.  In the
conformal limit this restriction amounts to the usual projection
of null states one encounters in the generalized Feigin-Fuchs
construction
\ref\rfl{G. Felder and A. LeClair, `Restricted Quantum Affine
Symmetry of
Perturbed Minimal Conformal Models', to appear in {\it Int. J. Mod.
Phys.}}.
For higher levels $k>1$, one must begin with a generalization of the
$\gh$-affine Toda theory so that it includes additional generalized
para-fermions for the group $G_k/ [U(1)]^{{\rm rank}(G)} $.  This
generalization has been called the fractional super-$\gh$-affine Toda
or (affine) para-Toda theory
\rfssg\rabl\ref\rnw{D. Nemeschansky and N. P. Warner,
USC preprint USC-91/031 (1991), `Topological Matter, Integrable
Models and Fusion Rings', to appear in { \it Nucl. Phys. B.}}; here
we will refer to the conformal, unperturbed combination of para-fermions
and free bosons as the para-Toda theory, and the perturbed,
integrable
model as the $k^{th}$ $\gh$-affine Toda theory or the $k^{th}$
affine para-Toda theory.  For $G=SU(2)$, this
latter model is the series of fractional super sine-Gordon models,
which
consists of an interacting system of a single boson and a $Z_k$
para-fermion
(for $k=2$ this is the usual supersymmetric sine-Gordon theory), and
is
described in detail in \rfssg .

In order to study general perturbations of $N=2$ theories, one must
understand
how to extend the foregoing results to the situation where $G\neq H$.
The
way to begin to accomplish this is contained in the paper
\rflmw\ for $k=1$ and for general $k$ in \ref\EHY{T.~Eguchi,
S.~Hosono and
S.-K.~Yang, \cmp{140} (1991) 159.}  \rnw. The main
observation made in these papers that we will use here is that the
perturbed
coset theories are again related to $\gh$-affine Toda theory, but now
with
a background charge tuned to $H$ rather than $G$.  In the sequel we
will
explain how to implement this non-conventional background charge in
the
restricted quantum group approach, and thereby derive the S-matrices
for
the perturbed $\co kl$ theories from the S-matrices of the $k^{th}$
$\gh$-affine Toda theories.
We will show how the conserved $N=2$ charges are always a subset of
the
$\uqGh$ quantum affine charges, with the $N=2$ supersymmetry occurring
at a special value of the coupling.

The remainder of this paper is organized as follows. In section 2, we
will review the relevant conformal field theories, and how their
perturbations are related to affine Toda field theory.  In section
3 we will describe in complete generality how one obtains the
S-matrices
from an appropriate restriction of the Toda theory.  Finally in
section 4
we will discuss specific examples in some detail.

\newsec{\bf The para-Toda models}

In this section we review the para-Toda construction of the coset
models:
\eqn\models{\mkl ~\equiv~ \Gkl{H} \ ,}
and their integrable perturbations.
Here $H$ is a subgroup of $G$ with
rank$(H)$ $=$ rank$(G)$\foot{All that is in fact
required  for the para-Toda construction is that $H$
is a regularly embedded subgroup of $G$.}.  As we will describe, the
$N=2$ superconformal coset models \rks\
are a subclass of these $\mkl$  models
\EHY
\rnw.
To fix our notation,
let $\alpha_1, \ldots , \alpha_r$ be a system of simple roots for
$G$,
ordered in such a way that $\alpha_1, \ldots , \alpha_p$, for $p \le
r$,
is a system of simple roots for $H$.  The highest root of $G$
will be denoted by $\psi$.
Let $\rho_G$ and $\rho_H$ denote the Weyl vectors, and $g$ and $h$
denote
the dual Coxeter numbers of $G$ and $H$ respectively \foot{Since $H$
is,
in general, a product of groups, $h$ is to be thought of as a vector.
The dual Coxeter number of a  $U(1)$ factor is defined to be zero.}.
Finally, let $\UU = (U(1))^r$
be a torus for $H$, and hence a torus for $G$.

Consider first the conformal field theory.
The para-Toda theory consists of the generalized para-fermions,
constructed
from the coset $G/\UU$, tensored with a model consisting of $r$ free
bosons.  The bosonic energy momentum tensor is
\eqn\em{
T_{b}(z) ~=~ - \half (\partial \phi )^2 ~+~ i ~(\beta_+^{(H)}
{}~-~ \beta_-^{(H)}) \rho_H \cdot \partial ^2 \phi \ ,}
where
\eqn\betadef{\beta_\pm^{(H)} ~\equiv~ {1 \over \sqrt{k}} ~ \Bigg[
\sqrt{{{(k+\ell + h)} \over {(\ell + h)}}}~ \Bigg]^{\pm 1}\ .}
As has been described in a number of places,  such a bosonic
free field description can be directly related to a Toda theory
(see, for example:
\ref\BilGe{A.~Bilal and  J.-L. \ Gervais, \plt{206} (1988) 412;
\nup{318} (1989) 579; \nup{326} (1989) 222;
A.\ Bilal, \nup{330} (1990)
399;  \ijmp{5} (1990) 1881.}
\ref\HM{T.~Hollowood and P.~Mansfield, \plt {226} (1989) 73.}
\rflmw ).

The primary fields of the para-fermionic theory will be denoted by
${\cal A}^\Lambda_\lambda (z)$, where $\Lambda$ is a highest weight
of
$G_k$, and $\lambda$ is a vector of charges under the Cartan
subalgebra
(CSA), ${\cal X}$, of $G$ that generates the torus, $\UU$.  The Toda,
or free bosonic,  field theory has a natural vertex operator
representation for its highest weight states:
\eqn\todavert{{\cal
V}_{\lambda_+, \lambda_-}(z) ~=~ exp \big[ - i (\beta_+^{(H)}
\lambda_+ ~-~ \beta_-^{(H)} \lambda_- ) \cdot \phi (z) \big ] \ .}
The conformal weight of ${\cal A}^\Lambda_\lambda$ is:
$$ h^\Lambda_\lambda ~=~ {{\Lambda
\cdot (\Lambda + 2 \rho_G ) } \over {2 (k + g)} }  ~-~ {{\lambda^2}
\over {2k}} ~+~ {\rm integer} \ ,$$
 and that of ${\cal V}_{\lambda_+, \lambda_-}$ is:
$$ h_{\lambda_+, \lambda_- }  ~=~ \half \big(
\beta_+^{(H)} \lambda_+ ~-~ \beta_-^{(H)} \lambda_- \big )^2 ~+~
(\beta_+^{(H)} ~-~ \beta_-^{(H)}) \rho_H \cdot (\beta_+^{(H)}
 \lambda_+ {}~-~ \beta_-^{(H)} \lambda_- ) \ .
$$
One finds that this can be rewritten as:
\eqn\confwt{h_{\lambda_+, \lambda_- }  ~=~  {1 \over {2k}}
{}~(\lambda_+
- \lambda_- )^2 ~+~ {{\lambda_+ \cdot (\lambda_+ + 2 \rho_H ) } \over
{2 (\ell + h)} } ~-~ {{\lambda_- \cdot (\lambda_- + 2 \rho_H ) }
\over
{2 (k + \ell + h)} } \ .}
{}From this, and a consideration of the
CSA eigenvalues, $\lambda$, it is easy to identify
representatives of the primary fields, $\primh$, of the $\mkl$ coset
model. ( The labels $(\Lambda, \lambda_+; \lambda_- )$ are highest
weight labels of affine $G$ at level $k$ and $H$ at levels
$\ell$ and $k+\ell$ respectively, and correspond to the numerator and
denominator factors in $\mkl$.)  Indeed, we may take
\eqn\primfield{\primh(z)~=~{\cal
A}^\Lambda_{(\lambda_- - \lambda_+)}(z)
{}~ {\cal V}_{\lambda_+,\lambda_-}(z)\ .}
One should also remember that there are field identifications
induced by spectral flow in the CSA, ${\cal
X}$, of $H$ \ref\MS{G.~Moore and  N.~Seiberg, \plt{220} (1989) 422.}
\ref\DGep{D.~Gepner, \plt{222} (1989) 207.} \ref\LVW{W.~Lerche,
C.~Vafa and N.P~Warner, \nup{324} (1989)427. }
\ref\SY{B.~Schellekens and S.~Yankielowicz, \nup{334} (1990) 67.}.
Such field identifications map a coset state
with weights $(\Lambda, \lambda_+ ;
\lambda_-)$ into another such state according to:
\eqn\flow {\Lambda ~\rightarrow~   \Lambda ~+~ k ~ v \ , \ \ \
\lambda_+ ~\rightarrow~  \lambda_+ ~+~ \ell  v\ , \ \ \ \lambda_{-}
 ~\rightarrow~   \lambda_{-} ~+~  (k+\ell) v \ , }
where $v$ is any vector.
Spectral flow by an arbitrary vector, $v$, yields an
automorphism of the coset theory provided we use appropriately
twisted
Kac-Moody currents (see, for example, \ref\GodOl{P. \ Goddard and D.
\ Olive, \ijmp{1} (1986) 303.}); that is, we replace the currents of
$G$ or $H$ according to:
$$\eqalign{J^{\, \alpha}_n ~\rightarrow~&
J^{\, \alpha}_{n + v\cdot \alpha} \cr H^i_n ~\rightarrow~& H^i_n ~+~
k
v^i \delta_{n,0}}$$
To avoid using such twisted representations one
usually restricts $v$ to be a weight of $G$, and hence a weight of
$H$.

Finally, to obtain the model $\mkl$ from the tensor product of the
para-fermions and free bosons one needs the screening currents,
and these are given by:
\eqn\screen{\eqalign{S_{+\alpha_i}(z) ~=~& \Phi^{0, -\alpha_i}_0 (z)
{}~\equiv~  {\cal A}^0_{\alpha_i}(z) ~ exp\big[ + i  \beta_+^{(H)}
{}~\alpha_i \cdot \phi (z) \big] \cr S_{-\alpha_i}(z) ~=~&
\Phi^{0, 0}_{- \alpha_i} (z)~\equiv~ {\cal A}^0_{- \alpha_i}(z)
{}~ exp\big[ - i \beta_- ~\alpha_i \cdot \phi(z) \big] \ ,}}
for $i = 1, \ldots, p$.

If $G/H$ is a symmetric space, one can conformally embed $H_{g-h}$
into  $SO({\rm dim}({G\over H}))$.  Therefore, for a special choice
of
modular invariant for $H_{g-h}$, the $M_{k,\ell=g-h}(G;H)$ model is
precisely the super-GKO coset model based on $G/H$.  That is,
one obtains the coset model:
\eqn\supcoset{{ G_k \times SO({\rm dim}(G/H)) \over H} \ . }
Moreover, if $G/H$ is a {\it hermitian}
symmetric space then $M_{k,\ell=g-h}(G;H)$ is an
$N=2$ supersymmetric model \rks.  For most of the remainder of this
section we will restrict our attention to the models  \models\ in
which $G/H$ is hermitian symmetric.  For such spaces the group $H$
has
the form $H=H^\prime \times U(1)$, where $H^\prime$ is semi-simple.
There is thus only one simple root, $\gamma \equiv \alpha_r$, of $G$
that is not a simple root of $H$.  The vector $2(\rho_G -\rho_H)$ has
the property that
\eqn\rhoprod{2(\rho_G ~-~ \rho_H) \cdot \alpha ~=~  \cases{ 0 &
\ \ if $\alpha $ is a root of $H^\prime$, \cr \pm g &\ \ otherwise;}}
where one has $\pm g$ depending upon whether $\alpha$ is a positive
or negative root.  Thus $2(\rho_G -\rho_H)$ defines the free $U(1)$
direction.

To determine exactly how to obtain
the $N=2$ model from the $\mkl$ model, one needs to decompose the
characters of  $SO({\rm dim}(G/H))$ into characters of $H_{g-h}$.
Let $\chi_R^{\pm}(\tau;\nu)$ and $\chi_{NS}^{\pm}(\tau;\nu)$ denote
the Ramond and Neveu-Schwarz characters of $SO({\rm dim}(G/H))$,
where
$\nu$ is the character parameter of the embedded CSA of $H$. One
finds that \rnw
\eqn\ramon{\chi_R^{\pm}(\tau;\nu) \ = \ \sum_{w \in {W(G) \over
W(H)}}
\sum_{\alpha \in {M(G) \over M(H)}} ~\epsilon^\pm (\lambda,w)
{}~\chi_{\lambda(\alpha,w)}^{H_{g-h}}(\tau ;\nu ) \ , }
where
\eqn\deflambda{\lambda(\alpha,w) \ = \ w(\rho_G) -
\rho_H + g \alpha \ , }
and
\eqn\defepsilon{\eqalign{ \epsilon^-(\lambda,w) \ & =
\ \epsilon (w) \ ,\cr  \epsilon^+(\lambda,w) \ & =
\ \epsilon (w) e^{-{2 \pi i \over g}  (\rho_G-\rho_H)\cdot
\lambda(\alpha,w)} \ , \cr }}
and $\chi^{H_{g-h}}_\lambda$ is the character  of $H$ at
level $g-h$ with highest weight $\lambda$.
The Weyl element, $w \in {W(G) \over W(H)}$, and the
vector $\alpha \in {M(G) \over M(H)}$ in \ramon\ are chosen so that
$\lambda(\alpha, w) $ is a highest weight of $H_{g-h}$.
The corresponding  result in the Neveu-Schwarz sector is almost
identical, except that \deflambda\ is replaced by
\eqn\nslabels{\lambda(\alpha,w) \ = \ w(\rho_G) -\rho_G +g \alpha \
.}
Note that the only difference between the sectors is a shift of
$\lambda(\alpha,w)$ by $\rho_G -\rho_H$, which is purely in the
$U(1)$
direction.  One should also observe that in the Neveu-Schwarz sector
$\lambda(\alpha,w)$ is always a {\it root} of $G$.

In the bosonic sector of the para-Toda formulation there is a single
free $U(1)$ direction that is orthogonal to the charge at infinity
and to the screening currents \screen.  This $U(1)$ factor is, of
course, the $U(1)$ current of the $N=2$ superconformal theory, and
is given by:
\eqn\superuone{J_{U(1)} (z)
{}~=~ 2i~ ( \beta_+^\prime ~-~ \beta_-^\prime ) ~ (\rho_G {}~-~
\rho_H)
\cdot \partial \phi \ ,. }
The associated charge, $q_{U(1)}$, of a primary field $\primh$ is:
\eqn\Qcharge{ q_{U(1)} ~=~ -2 (\rho_G ~-~
\rho_H) \cdot \bigg[ {{\lambda_+} \over g} {}~-~ {{\lambda_-}
\over {k + g}} \bigg ] \ .}
Recall that in the Neveu-Schwarz sector of the $N=2$ model, the
vector
$\lambda_+$ is a root of $G$ and hence
${2 \over g} (\rho_G -\rho_H)\cdot \lambda_+$ is always an integer.
The same conclusion is also true in the Ramond sector since
$2(\rho_G -\rho_H)^2$ is also always an integer.

The para-Toda formulation is not manifestly supersymmetric, but the
supercharges have natural realizations in terms of vertex
operators.  Indeed, the operators
\eqn\susys{\eqalign{S_{\gamma}(z) ~=~& \Phi^{0, -\gamma}_0 (z)
{}~\equiv~  {\cal A}^0_{\gamma}(z) ~ exp\big[ + i  \beta_+^{(H)}
{}~\gamma \cdot \phi (z) \big] \cr S_{-\psi}(z) ~=~&
\Phi^{0, \psi}_0 (z)  ~\equiv~  {\cal A}^0_{-\psi}(z)
{}~ exp\big[ - i  \beta_+^{(H)}  ~\psi \cdot \phi (z) \big]
 \ }}
are representations of $G^+(z)$ and $G^-(z)$ respectively.
The symmetric relationship between these operators and the
screening currents has important implications for the
structure of the super-$W$ algebra \ref\LNW{W.~Lerche,
D.~Nemeschansky and N.P.~Warner, in preparation.}.

There are two other operators that are closely related
to the screening currents:
\eqn\perts{\eqalign{ S_{-\gamma}(z) ~=~&
\Phi^{0, 0}_{- \gamma} (z)~\equiv~ {\cal A}^0_{- \gamma}(z)
{}~ exp\big[ - i \beta_-^{(H)} ~\gamma \cdot \phi(z) \big] \ ,
\cr S_{\psi}(z) ~=~&  \Phi^{0, 0}_{\psi} (z)~\equiv~
{\cal A}^0_{\psi}(z)  ~ exp\big[ + i \beta_-^{(H)} \psi \cdot
\phi(z) \big] \ .}}
These are representations of $(\Gminus X)(z)$ and
$(\Gplus \tilde X )(z)$, where $X(z)$ is the chiral, primary
field
\eqn\cppert{X ~=~ \Phi^{0, -\gamma}_{-\gamma}\ ,}
with $h =
{ k \over {2(k+g)}}$ and $q_{U(1)} = { k \over {(k+g)}}$.  The field
$\tilde X (z)$ is the anti-chiral conjugate of $X(z)$.

We now turn to massive integrable perturbations.
The set $\{ \al_0 , .., \al_r \}$, with $\al_0 = -\psi$, comprise
the simple roots of the affine Lie algebra $\hat{G}$.
Define the $k^{th}$
$\hat{G}$-affine Toda theory at coupling $\beta$ by the action:
\eqn\action{
S = \inv{4\pi} \int d^2 z ~ \d_z \Phi \cdot \d_{\zb} \Phi
+ S_{ G_k /\CU }
+ \frac{\la}{2\pi} \int d^2 z  \sum_{i=0}^r
\CA^{(0)}_{\al_i} {\bar{\CA}}^{(0)}_{\al_i } \>
\exp \( -i \be \al_i \cdot \Phi \) ,}
where $S_{G_k /\CU}$ is the formal action of the para-fermions and
 the operator ${\bar{\CA}}^{(0)}_{\al_i }$ is the anti-holomorphic
counterpart of ${{\CA}}^{(0)}_{\al_i }$.
The terms with  $i=1,\ldots ,p$ in the sum in \action\ characterize
the
$\CM_{k,l} (G;H)$ conformal field theory, when $\beta =
\beta^{(H)}_-$.
The additional terms ($i=0$ and $i=p+1,\ldots,r$) are to be thought
of as  perturbations of the conformal field theory.  The fact that
\action\ is a classically integrable model suggests very strongly
that
the model will be quantum integrable.  From the perspective of
perturbed conformal field theory, the perturbing operators in
\action\ are related to the screening operators
 by an automorphism of the Lie algebra of $G$.  This provides
very strong evidence that the perturbed
conformal model is indeed a  massive, quantum integrable
field theory \ref\EY{ R. Sasaki and I. Yamanaka, in
`Conformal Field Theory and Solvable Models',
{\it Advanced Studies in  Pure Mathematics } {\bf 16}, 1988.
T.~Eguchi and S-K. Yang, Phys. Lett. {\bf B 224} (1989) 373;
{\it Mod. \ Phys. \ Lett.} {\bf A5} (1990) 1693; } \rflmw.
Note that for the $N=2$ supersymmetric models, the perturbations
in \action\ are simply the operators \perts\ (paired, of course,
with the anti-holomorphic counterparts).  When $G=H$ there is
only one perturbing operator (the $i=0$ term in \action) and it
corresponds to the field  $\Phi^{0,0}_{\rm Adj}$ in $G \times G/G$,
where ${\rm Adj}$ refers to the adjoint representation of $G$.
In the situation where $G$ has level one ({\it i.e.} $k=1$),
this way of relating integrable perturbations of conformal theories
to affine Toda theory was described in \EY\HM\rflmw
\ref\SMNW{S.~Mathur and N.P.~Warner, \plt{254} (1991) 365.}.

If we start from an $N=2$ superconformal field theory then
the massive perturbed model is also  $N=2$ supersymmetric.
In addition, if the degeneracy of the
Ramond ground state of the conformal model is $\mu$, then
the perturbed theory has $\mu$ distinct ground states that are
fully resolved by the different expectation values of the
chiral primary fields in those ground states
\rnw.  Moreover, if $G$ has level one, then
there is a \LG formulation and there are $\mu$ distinct minima
of the potential.   All the classical small oscillations about these
minima correspond to massive excitations
\LVW\rlg.
  This sort of vacuum structure leads one to
consider the soliton sector of the effective field theory for the
chiral primary fields.  A great deal can be said about the
soliton structure\rntwo\rlg,
but for the present we simply wish to
note that if one introduces the soliton sectors into the
perturbed conformal model and then returns to a conformal model
by taking the ultra-violet limit then one has gone beyond the usual
(modular invariant) conformal model.  This is because, in the
ultra-violet limit,  the  soliton creation operators become purely
holomorphic or purely anti-holomorphic fields with possibly
fractional
fermion number.  A super-selection rule was advanced in
\rfl\ to determine which operators in
the conformal field theory correspond to the soliton operators
of the perturbed model. This selection rule states that one
should include all the holomorphic and anti-holomorphic
operators that are local with respect to the perturbing operators.

Consider once again the completely general conformal $\mkl$ model.
It is elementary to identify the operators that are local with
respect to the perturbing operators \perts.  Since these perturbing
operators have the form $\Phi_{\alpha}^{0,0}(z)$, it is clear that
any operator of the form $\Phi_{0}^{\Lambda, \lambda_+}(z)$ is local
with respect to the perturbation.  The fact that one has
$\lambda_- = 0$ means that $\Lambda + \lambda_+$ must be on the root
lattice  \foot{  This lattice is of dimension $r$ with
appropriate basis vectors chosen in the $U(1)$ directions.  When
$G/H$ is a hermitian symmetric space this lattice has basis vectors
$\alpha_1, \ldots, \alpha_{r-1}, 2(\rho_G - \rho_H)$.} of $H$.
If $G$ has level one then $\Lambda$ is uniquely determined by
the choice of $\lambda_+$.   The problem now is
to determine what subset of the
operators $\Phi_{0}^{\Lambda, \lambda_+}(z)$ correspond to soliton
creation operators, and what are the set of possible choices for
$\lambda_+$.  For the $N=2$ superconformal models it is
tempting to restrict $\lambda_+$ to those of \ramon\ and
\nslabels, however the resulting operators will only have
integer fermion number, $q_{U(1)}$, while the work of
\rfi\ shows that solitons
can have fractional fermion number.  In the next section
we obtain a consistent scattering matrix for the solitons of
the abovementioned integrable models, and our results lead to the
conclusion that the vectors, $\lambda_+$, in the soliton creation
operators can be any of the fundamental weights of $G$.  In this
way we will also recover the results of \rfi.

Before concluding this section, we wish to remark upon the
generalizations beyond the $N=2$ supersymmetric integrable
models.  For general $G$ and $H$, with rank($G$) $=$ rank($H$),
one can certainly construct the conformal model \models\ using
the para-Toda formalism, however if one wants to construct a
unitary integrable model by using perturbations similar to
\perts\ in an action of the form \action,
then $G/H$ must be a symmetric space \SMNW \foot{ Note that we
define the trivial quotient $G/G$ to be symmetric.}.
If $G/H$ is a real symmetric space then there is only one
operator that is analogous to \perts\ and which leads to an
integrable model.  This operator is real and so
perturbation hamiltonian is hermitian.  If $G/H$ is a hermitian
symmetric space then the two perturbations \perts\ are hermitian
conjugates of each other.  If $G/H$ is not symmetric then there
is no way to use operators of the form \perts\ to obtain a hermitian
perturbation hamiltonian:  The extra terms, $i=0$ and
$i =p+1, \ldots, r $ in \action\ do not possess any form of
hermiticity in the quantum group truncated model.  Thus the
restriction to symmetric spaces, though not necessarily to hermitian
symmetric spaces, is required.  The restriction
to supersymmetric models is optional and would require
$\ell = g-h$ and a special choice of modular invariant.  Therefore
the freedom to generalize the quantum integrable models arising out
of perturbations of the $\mkl$ models is not so much in the choice of
$H$ as it is in the choice of $\ell$ and the modular invariant for
$\mkl$.

\newsec{\bf The $\uqH$ restriction of the $\gh$-affine Toda theories
}
\medskip

Let us summarize the results of the previous section.  The perturbed
$\co kl$ models were shown to be related to the $k^{th}$ $\gh$-affine
Toda field theory at the specific coupling $\beta = \beta^{(H)}_-$.
The energy-momentum tensor in the conformal limit has the background
charge appropriate to $H$.
In this section we describe how to obtain the spectrum and S-matrices
of the perturbed coset theories from the S-matrices of the
$k^{th}$ $\gh$-affine Toda theory.  We first review the S-matrices
of the $\gh$-affine Toda theory with zero background charge, and then
describe how to restrict the model in a way appropriate to the
value of the background charge \em .

For simplicity we begin with the case $k=1$.  The S-matrices for the
$\gh$-affine Toda theory (at `imaginary coupling', which is the only
case of relevance here) can be determined by requiring that they
commute
with the $\uqGh$ quantum affine symmetry that exists in the model
\rnlc .
Let us review this construction. Let $\av i , i= 0,..,r$ be the
simple
roots for the affine Lie algebra $\gh$.
For simplicity we assume that  $\gh$ is simply laced \foot{The
following results can be generalized to the non-simply laced case
\rfl .}
so that $\av i ^2 = 2 $.  The $\gh$-affine Toda theory with the
coupling $\be$ is defined by the action
\eqn\eIIiii{
S = \inv{4\pi} \int d^2 z ~ \d_z \Phi \cdot \d_{\zb} \Phi
+ \frac{\la}{2\pi} \int d^2 z  \sum_{i=0}^r \exp \( -i \be \al_i
\cdot \Phi \)
.}
Let $\phi$ and $\phib$ denote the non-local quasi-chiral components
of the
field $\Phi = \phi + \phib$:
$$ \phi (x,t) = \inv{2} \( \Phi (x,t) + \int_{-\infty}^x  \ dy
\d_t \Phi (y,t) \) $$
$$ \phib (x,t) = \inv{2} \( \Phi (x,t) - \int_{-\infty}^x  \ dy
\d_t \Phi (y,t) \) .  $$
One can show that the model \eIIiii\ possesses
non-local conserved charges $\Q {\al_i} , \Qb {-\al_i} , i=0,..,r$
which
together with the topological charges
\eqn\eIIiv{
h_i = \frac{\be}{2\pi} \int dx ~ \al_i \cdot \d_x \Phi , }
generate the $\uqGh$ quantum affine algebra.  The conserved currents
$J^\mu_{\al_i} , \bar{J}^\mu_{-\al_i}$ for the charges $\Q {\al_i} ,
\Qb {-\al_i}$ respectively, which satisfy $\d_\mu J^\mu_{\al_i}
= \d_\mu \bar{J} ^\mu_{-\al_i } = 0$, are the following:
\eqn\eIIv{\eqalign{
J_{\al_i ,z} (z,\zb ) &= \exp \( \frac{i}{\be} \al_i \cdot \phi \)
\cr
J_{\al_i ,\zb } (z,\zb ) &= \la \frac{\be^2}{1-\be^2}
\exp \( -i (\be - \inv{\be} ) \al_i \cdot \phi - i \be \al_i \cdot
\phib \)
\cr
\bar{J}_{-\al_i ,\zb} (z,\zb ) &= \exp \( \frac{i}{\be} \al_i \cdot
\phib \)
\cr
\bar{J}_{-\al_i ,z} (z,\zb ) &= \la \frac{\be^2}{1-\be^2}
\exp \( -i (\be - \inv{\be} ) \al_i \cdot \phib - i \be \al_i \cdot
\phi \)
\cr. }}

These charges can be shown to satisfy the $\uqGh$ algebra:\foot{The
following relations are isomorphic to the standard relations in
\ref\rJim{M. Jimbo, \lmp{10} (1985) 63; \lmp{11}  (1986) 247.}
\ref\rdrin{V. G. Drinfel'd, \smd{32} (1985) 254;
\smd{36} (1988) 212.} . }
\eqn\eIIvi{\eqalign{
[h_i , \Q {\al_j} ] = a_{ij} Q_{\al_j} , ~~&~~~
[h_i , \Qb {-\al_j} ] = -a_{ij} \Qb {-\al_j} \cr
\Q {\al_i} \Qb {-\al_j } - q^{-a_{ij} } \Qb {-\al_j} \Q {\al_i}
&= a \de_{ij} \( 1- q^{2h_i} \) , \cr }}
where $a_{ij} = \al_i \cdot \al_j $ is the Cartan matrix, $a$ is a
constant, and $q$ is given by
\eqn\eIIvii{
q= \exp \(-\frac{i\pi}{\be^2} \) ~~~~~~~~~~~~(k=1). }
The deformed Serre relations were proven in \rfl .
The Lorentz spin of the quantum affine charges is given by
\eqn\eIIviib{
\inv{\gamma} \equiv {\rm spin} (Q_{\al_i} ) = - {\rm spin} (\Qb
{-\al_i} )
	       = \frac{1-\be^2}{\be^2} . }

We now describe the spectrum of massive particles of the
$\gh$-affine Toda field theory.  As usual we parameterize
the momentum of asymptotic one-particle states in terms of
rapidity $\th$,
\eqn\eIIviii{
E=m\cosh (\th) , ~~~~~P = m\sinh (\th) . }
The spectrum consists of $r$ separate finite dimensional multiplets
of solitons $\k n (\th ), n=1,..,r$.  The solitons in each separate
multiplet $\k n $ are all of the same mass $M_n$, and are
characterized as transforming under a representation $\rh n$ of
$\uqGh$.  We denote by $\w n$ the finite dimensional representation
vector space of $\rh n$, such that the states $\k n $ are
vectors in $\w n$.

The representation vector spaces $\w n$ are described more precisely
as follows.  Let $\v n$, $n=1,\ldots,r$ denote the vector spaces of
the  fundamental representations of $G$.  For each of the fundamental
representations $\v n$ that correspond to integrable representations
in the $G$-Wess-Zumino-Witten (WZW) model at level one (the number of
these is always less than or equal to $r$), there
exists a multiplet of solitons where $\w n = \v n$.  The fields that
create these solitons can be taken as
\eqn\eIIix{
K^{(n)}_{\mu_n } (z,\zb ) = \exp \( - \frac{i}{\be} \mu_n \cdot
\phi (z,\zb ) \) , ~~~~~{\bar{K}}^{(n)}_{\mu_n} (z,\zb ) =
\exp \( \frac{i}{\be} \mu_n \cdot \bar{\phi}  (z , \zb ) \) , }
where
$\mu_n$ is any weight in $V_n$.  The fields \eIIix\ are local
with respect to the perturbing field, and   thus define meaningful
superselection sectors\rfl .
For example for $G=SU(N)$, $\w n = \v n$ for all $n$.  For other
groups
the remaining $\w n$ are direct sums of a finite number of $V_n$, and
as vector spaces are not reducible with respect to $G$. The
precise decomposition of the spaces $\w n$, as well as the masses
$M_n$ are known for all
$G$\ref\rkath{\KaTh}\ref\row{\OW}\ref\rorw{\ORW},
and can be traced to the representation theory of the algebra
$\uqGh$\foot{
The papers \row\rorw\ are not concerned with $\gh$-affine Toda
theory, but
rather with the generalized Gross-Neveu models, or non-abelian
Thirring
models, which actually have a G-Yangian invariance.  One can appeal
to
these results in the present context since the $\gh$-affine Toda
theories
are equivalent to the Yangian invariant ones as $\be \to 1$.  We
refer
the reader to \rnlc\ for details of this argument.}.

The representations $\rh n$ of $\uqGh$ are rapidity dependent due to
the non-zero Lorentz spin of the charges.  This can be expressed as
follows:
\eqn\eIIx{
\rh n \( Q_{\al_i} \) = x \> {\rh n}^\prime \( Q_{\al_i } \) ,
{}~~~ ~~ \rh n \( \Qb {-\al_i } \) = x^{-1}\>  {\rh n}^\prime \( \Qb
{-\al_i }
\) , }
where
\eqn\eIIxi{
x \equiv \exp (\th /\gamma ) }
is a `spectral' parameter, and ${\rh n}^\prime $ are rapidity
independent
representations of $\uqGh$.  The $x$ dependence of the representation
\eIIxi\ defines it to be in the so-called principal gradation.

Define $S^{(\gh )}_{nm} (\th , q)$ to be the two-body
S-matrix for the scattering
of particles in $W_n$ with those in $W_m$:
\eqn\eIIxii{
S^{(\gh )}_{nm} (\th, q) : ~~~~~W_n \ot W_m \to W_m \ot W_n   }
$(\th = \th_1 - \th_2 )$.
Requiring the S-matrices to commute with the $\uqGh$ symmetry leads
to the
following result:
\eqn\eIIxiii{
S^{(\gh )}_{nm} (\th , q) = X_{nm} (\th )\>  v_{nm} (\th, q) \>
R^{(\gh )}_{nm} (\th , q ). }
In \eIIxiii\ the various factors have the following meaning.
The last term,
$R^{(\gh )}_{nm} (\th , q) $, is the standard $R$-matrix for the
quantum group $\uqGh$ in the principal gradation, and its structure
is completely fixed by the $\uqGh$ symmetry.  It also automatically
satisfies the
Yang-Baxter equation.  The rapidity dependence enters through the
spectral parameter $x$, whereas the coupling $\be$ dependence enters
through both $x$ and $q$.  Many of the $R^{(\gh )}_{nm}$ are
explicitly
known\ref\rJimcmp{\Jimboii}\ref\rbazan{V. V. Bazhanov, \cmp{113}
 (1987) 471.};  they can be computed in principle from the fusion
procedure\ref\rkrs{\fusion}.  See \ref\rjimborev{\Jimborev}\ for a
review
\foot{One needs to be careful about the rapidity dependence of the
spectral parameter $x$ when borrowing $R$-matrices from the
mathematics literature. For example, in \rJimcmp\ the $R$-matrices
by construction commute  with the affine charges in the homogeneous
gradation, and are related to the $R$-matrices in \eIIxiii\ by
an automorphism (see below).}.
The scalar factor $v_{nm} (\th , q)$ is the minimal
factor that makes the product $v_{nm} R_{nm}$ crossing symmetric
and unitary. For $G=SU(N)$ they were computed in \ref\rhol{\Hollow}.
For more general groups they are straightforward, though tedious,
to compute.  The additional factor $X_{nm} (\th )$ is independent of
the coupling $\be$, and is a CDD factor.  This factor contains all
of the necessary poles for closure of the bootstrap with the
spectrum of masses $M_n$.  The $X_{nm}$ are also known from the
relation with the Gross-Neveu type models\row\rorw .

We now turn on the background charge \em, and restrict the model
to obtain the S-matrices for the perturbed cosets.  This proceeds
in two stages.  One must first modify the S-matrices \eIIxiii\ such
that they reflect the presence of the background charge.  One then
uses the screening quantum group sub-algebra to truncate the
Hilbert space.

The non-zero background charge modifies the conformal dimension
of the quantum affine currents \eIIv.  A simple computation shows
that the charges $Q_{\al_i}, \Qb {-\al_i }$, for $\al_i$ a root of
$H$ become dimension zero operators, and thus have Lorentz spin zero;
these generate a screening quantum group sub-algebra $\uqH$.
The remaining quantum affine charges have modified non-zero Lorentz
spin
(which is easily computed from in the conformal limit from \eIIv\ and
\em )
and will generate residual quantum symmetries of the S-matrices of
the
kind described in \ref\rzamoiv{\Zamoiv}\ref\rBLii{\BLii}\rfssg\rfl .
The effect of the background charges on the representations $\rh n$
can be expressed as follows.   Let
${\rh n}^{(H)}  \( Q_{\al_i} \)$, $
{\rh n}^{(H)} \( \Qb {-\al_i } \)$
denote the representations of $\uqGh$ when the background charge is
present.  Then $\rh n$  and $\rhh n$ are related by an automorphism:
\eqn\eIIxiv{
{\rh n}^{(H)}  \( Q_{\al_i} \)
= \sigma_H^{-1} \> {\rh n} \( Q_{\al_i} \) \> \sigma_H, ~~~~~
{\rh n}^{(H)} \( \Qb {-\al_i } \)
= \sigma_H^{-1} \>  {\rh n} \( \Qb {-\al_i } \) \> \sigma_H , }
where
\eqn\eIIxivb{ \sigma_H = x^{\rho_H \cdot h }  }
and  $h$ in the foregoing exponent represents a vector of Cartan
elements.   When $G=H$, the representations $\rh n ^{(G)}$ are
sometimes referred to as being in the homogeneous gradation.

Since the S-matrices are completely characterized by their
$\uqGh$ symmetry, one can deduce the effect of the background
charge on the S-matrices from \eIIxiv . Let $S^{(\gh /H)}_{nm}$
denote such an S-matrix.  It is given by the formula
\eqn\eIIxv{
S^{(\gh /H)}_{nm} (\th , q) =
\( \sigma_H^{-1} \ot \sigma_H^{-1} \) \>
S^{(\gh ) }_{nm} (\th , q) \>
\( \sigma_H \ot \sigma_H \) . }

\def\kn#1#2{{ K^{(#1)}_#2 }}

By construction, the S-matrices $S^{(\gh /H)}_{nm}$ commute with the
action of the finite quantum group $\uqH$. These charges act in
a rapidity independent fashion on the states, since they have Lorentz
spin zero.  Therefore one can use the $\uqH$ symmetry to restrict the
model.  The restriction may be described as follows.  For each weight
$\alpha$ of $W_n$ we introduce formal operators
$\kn n \al (\th )$.   These operators
satisfy an S-matrix exchange relation:
\eqn\eIIxvi{
\kn n \al (\th_1 ) \> \kn m \be (\th_2 )
= \sum_{\al ', \be '}
\( S^{(\gh /H)}_{nm} (\th , q) \)^{\al ' \be '}_{\al \be} ~
K^{(m)}_{\be '}  (\th_2 ) \> K^{(n)}_{\al '}  (\th_1 ) . }
Let $\CF$ denote the multiparticle fock space generated by the formal
action of the operators $\kn n\al (\th )$ on the vacuum. The space
$\CF$ is an $\uqH$ module, and reducible (for $q$ not a root of
unity):
\eqn\eIIxvii{ \CF = \bigoplus_i  V^{( \lhi i )} , }
where
$V^{(\lh )} $ is an $\uqH$ module of highest weight $\lh$.
Since the $\kn n\al (\th )$ act on $\CF$, one can consider their
reduction
\eqn\eIIxviii{
K^{(n)}_{\lhi j \lhi i } (\th ) : ~~~~~ V^{(\lhi i )} \longrightarrow
V^{(\lhi j )} . }
These operators satisfy the exchange relation:
\eqn\eIIxix{\eqalign{&
K^{(n)}_{\lhi j \lhi i } (\th_1 )\>
K^{(m)}_{\lhi i \lhi k } (\th_2 )
= \cr & \sum_{\lhi l}
\( S^{(\gh /H)}_{nm} (\th , q) \)^{\lhi j \lhi l}_{\lhi i \lhi k} ~
K^{(m)}_{\lhi j \lhi l } (\th_2 )\>
K^{(n)}_{\lhi l \lhi k } (\th_1 )  . \cr}}
The S-matrix for the kinks in \eIIxix\ is in the so-called SOS
(solid-on-solid) form.  The above construction is the usual
vertex/SOS correspondence which is commonplace in lattice
statistical mechanics\ref\rpas{V. Pasquier, Commun. Math. Phys.
{\bf 118}
(1988)
355.}.

Finally, the restriction amounts to taking $q$ to be a root of unity
and
imposing a limitation on the allowed highest weight labels $\lh$,
due to the fact that $q$ is a root of unity.  From \betadef\ and
\eIIvii\ one finds that $q$ has the value
\eqn\eIIxx{ q= - \exp \( -i\pi /(l+h) \).}
The limitations on the labels $\lh$ can be deduced from the
representation theory of $\uqH$ \ref\rpassal{\PasS}; the result is
equivalent to the statement that $\lh$ must correspond to an
integrable
representation of the $H$-$WZW$ model at level $l$.  Additionally,
the
pair $\lhi j , \lhi i$ must be admissable, which means that
$V^{(\lhi j )}$ must be contained in the tensor product
$W_n \ot V^{(\lhi i )}$ when the spaces are considered as
$H$ modules.

To summarize, the spectrum of the perturbed $G_1 \ot H_l/H_{l+1}$
coset theory consists of RSOS kinks $K^{(n)}_{\lhi j \lhi i} (\th )$,
with the above limitations on the labels $\lh$, characterized by
an integer $l$.  The S-matrix for these kinks is the RSOS form
of  $S^{(\gh /H)}_{nm}$
which we denote as
 $S^{(\gh /H);l}_{nm}$.
For $G=H$ the above result was described in \rabl\ref\rnak{\Nak}
\ref\rvefa{\VeFa}.

The action of the residual symmetries on the kink states is
described as follows.  The residual charges are
$Q_{\al} , \bar{Q}_{-\al}$ for $\al$ not a root of $H$.  Each of
these charges can be associated to a representation $\lambda^{(H)}_{\al}$
of $H$ by considering $\al$ a weight of $H$.  The charges must
be decomposed into the components $\( Q_\al \)_{\lambda^{(H)}_2
\lambda^{(H)}_1 }$
that intertwine the sectors $\lambda^{(H)}_2$ and $\lambda^{(H)}_1$, since
these are what have a well-defined action on the states.  The currents
for these intertwiners  have well-defined, generally
non-abelian, braiding relations.
One finds
\eqn\esec{
\( Q_\al \)_{\lambda^{(H)}_3 \lambda^{(H)}_2 } ~
| K^{(n)}_{\lambda^{(H)}_2 \lambda^{(H)}_1 } (\th ) \rangle
= e^{s \th} ~
\left\{ \matrix{\lambda^{(H)}_3 & \lambda^{(H)}_\al & \lambda^{(H)}_2 \cr
\lambda^{(H)}_n & \lambda^{(H)}_1 & \lambda^{(H)}_n \cr}\right\}_q
{}~
| K^{(n)}_{\lambda^{(H)}_3 \lambda^{(H)}_1} (\th ) \rangle , }
where $s$ is the Lorentz spin of $Q_\al$,
$\{ * \}_q $ represents a generalized $6j$ symbol for
$U_q (H)$, and $\lambda^{(H)}_n$ refers to the space $W_n$ viewed
as a $U_q (H)$ module.  Similar formulas give the action of
$\bar{Q}_\al$ with $s\to -s$.  One has the qualitative rules
\eqn\equal{
\( Q_\al \)_{\lambda^{(H)}_3 \lambda^{(H)}_2 }~
| K^{(n)}_{\lambda^{(H)}_2 \lambda^{(H)}_1 } (\th ) \rangle
{}~ \neq ~ 0 ~~~~~~{\rm if} ~~V^{\lambda^{(H)}_3} \subset
V^{\lambda^{(H)}_\al}  \otimes V^{\lambda^{(H)}_2 } . }
General quantum group theoretic arguments show that this action is
necessarily a symmetry of the above S-matrix.
For $G=H=SU(2)$ this construction was described in detail in \rBLii\rfl.

For $G=H$ the above result was described in \rabl\ref\rnak{\Nak}
\rvefa.
The remaining quantum affine charges in this situation are
$Q_{\al_0}$ and $\Qb {-\al_0}$, with fractional spin $\pm g/(g+l)$,
and they generate residual
quantum symmetries of the RSOS S-matrices. In order to describe
properly the action of these charges on the RSOS kinks and verify
that they commute with the S-matrices $S^{(\gh /G);l}$ one must
screen them in a manner described for $SU(2)$ in \rfl\rBLii .
In a coset description these symmetries can be identified with
the fractional supersymmetries generated by the conserved currents
\eqn\eIIxxi{
J^{(l)} =   \Phi^{0,{\rm Adj}}_0  }
(for $k=1$).
We will use this fact later.

Let us specialize now to the perturbations of the $N=2$ theories,
again when $k=1$.  In this situation, the level $l=g-h$, and
$q=\exp (-i\pi /g )$.  Furthermore $H=H' \ot U(1)$, so that
$\rho_H = \rho_{{H'}}$.  Therefore the restriction described above
is performed with the quantum group $\uqHp$.  The spectrum consists
of RSOS kinks $K^{(n)}_{\lambda^{({H} )}_j \lambda^{({H} )}_i }
(\th )$, whose scattering is given by the S-matrices
$S^{(\gh /{{H'}} ) ; g-h }_{nm}$.  When the background charge is
as in \em, the quantum affine charges $Q_{\al_0} , \Qb {-\al_0}$
and $Q_{\al_r} , \Qb {-\al_r}$ have Lorentz spin $\pm 1/2$, and with
proper screening are identified with the $N=2$ supercharges.
Due to the automorphisms $\sigma_H$ that define $S^{(\gh /{{H'}}
);g-h }$,
by design this S-matrix commutes with the action of these
supercharges.

The $U(1)$ current of the $N=2$ superconformal algebra is given by
\eqn\eIIxxii{
J_{U(1)} (z) = \frac{2i}{\sqrt{g(g+1)}} (\rho_G - \rho_H )\cdot \d
\phi ,
}
and similarly for $\bar{J}_{U(1)}$.  These currents are normalized such that
the $U(1)$ charge of the $N=2$ supercharges is $\pm 1$.   The
perturbed
theory is invariant under a diagonal $U(1)$.  The $U(1)$ charges of
the
kink states can be computed from the vertex operators \eIIix\ that
create them, at least when $W_n = V_n$.  For the kinks
$K^{(n)}_{\mu_n}$ one obtains
\eqn\eIIxxiii{
q_{U(1)} = - \inv{g} ~ 2(\rho_G -\rho_H ) \cdot \mu_n . }
For example, for $G=SU(N+1)$, ${H'} = SU(N)$, these $U(1)$ charges
are multiples of $1/(N+1)$, in accordance with the fractional fermion
numbers
in \rfi\rfii .

Consider now the case of level $k>1$.  The arguments leading to the
spectrum and S-matrices given in \rfssg\rabl\ for the case $G=H$ can
be
repeated with little modification.  We briefly outline the argument
and
the result.  The $k^{th}$ $\gh$-affine Toda theory is characterized
by
a $\uqGh$ symmetry for all $k$.  The $\uqGh$ currents are simple
modifications
of the currents in \eIIv , where now the quasi-chiral components
$J_{\al_i ,z}$ and $\bar{J}_{-\al_i ,\zb}$ of the currents are
multiplied
by $G_k/[U(1)]^r$ para-fermions $\CA^{(0)}_{\al_i} ,
\bar{\CA} ^{(0)}_{-\al_i} $
respectively:
\eqn\epara{\eqalign{
J_{\al_i ,z} (z,\zb ) &= \CA^{(0)}_{\al_i } (z , \zb ) \>
\exp \( \frac{i}{\be k } \al_i \cdot \phi \)  \cr
\bar{J}_{-\al_i ,\zb} (z,\zb ) &= \bar{\CA}^{(0)}_{-\al_i } (z , \zb
) \>
\exp \( \frac{i}{\be k } \al_i \cdot \phib \)  . \cr }}
The other components of the conserved currents can be deduced from
conformal perturbation theory. The spin of the para-fermions
$\CA^{(0)}_{\al_i} , \bar{\CA}^{(0)}_{-\al_i }$ is
$\pm (k-1)/k $.  Therefore the spin of the $\uqGh$ charges now
becomes
\eqn\espin{
\inv{\gamma_k} \equiv {\rm spin} (Q_{\al_i} ) = - {\rm spin}
( \Qb {-\al_i} ) = \frac{1-k\be^2}{\be^2 k^2} . }
The braiding of the para-fermions, in addition to the braiding of
the vertex operators in the currents \epara, now imply that
$q$ is changed to $-\exp (-i\pi /{\gamma_k} )$.  Therefore the
representations \eIIx\ are valid with
\eqn\enew{
x= \exp (\th / {\gamma_k} )~~, ~~~~~~~q= - \exp ( -i\pi / {\gamma_k}
). }

For $k>1$, the $k^{th}$ $\gh$-affine Toda theory has
some additional symmetries generated by
charges $Q^{(k)}, \qb^{(k)}$, not present at $k=1$.  These symmetries
have fractional spin $\pm g/(g+k)$, and are independent of the
$\uqGh$
symmetries. In the conformal limit,
the currents for these symmetries are of the form
$\epsilon \> \d \phi$, where $\epsilon$ is an `energy' operator in
the
para-fermion theory.  When the background charges are turned on,
in the coset theory this current corresponds to
the field
\eqn\eIIxxiv{
J^{(k)} =  \Phi^{ {\rm Adj},0}_0  .  }
In the conformal theory, these currents play the role of generating
a non-local chiral algebra for the coset models
\ref\rcosetff{\cosetFF}.
In the massive theory these symmetries are unbroken and must be
symmetries of the S-matrix.
The fact that the S-matrix must be invariant under two independent
sets of symmetries $Q^{(k)}, \qb^{(k)}$ and $\uqGh$ implies that
it must be the tensor product of two factors.  The $Q^{(k)} ,
\qb^{(k)}$
invariant factor can be deduced as follows.
When $G=H$, under the  $k\Leftrightarrow l$ duality of the
coset models, the $Q^{(k)} , \qb^{(k)}$
symmetries are dual to the residual $\uqGh$ symmetries coming from
the
current \eIIxxi;  thus one concludes that the S-matrix contains a
factor $S^{(\gh /G) ;k }$.
The spectrum of the $k^{th}$ $\gh$-affine Toda theory thus consists
of
kinks with a RSOS $\ot$ vertex structure $K^{(n)}_{ \lambda^{(G)}_j
\lambda^{(G)}_i; \al}$ of mass $M_n$, where $\al$ is a weight of
$W_n$.
The S-matrix is given by
\eqn\etoda{ S^{k^{th} \gh{\rm -Toda}}_{nm} (\th ,q)
= X_{nm} (\th ) ~  \tilde{S}^{(\gh /G);k }_{nm} (\th ) \> \ot \>
\tilde{S}^{(\gh )}_{nm} (x, q) , }
where
$\tilde{S}_{nm} \equiv S_{nm} /X_{nm}$.  The factor $\tilde{S}^{(\gh
)}$
acts on the $W_n$ indices of the kinks and is $\uqGh$ symmetric,
whereas
$\tilde{S}^{(\gh /G);k}$ acts on the $(\lambda^{(G)}_j
\lambda^{(G)}_i)$
RSOS labels.

When the background charges are turned on, only the $\uqGh$
symmetries
are affected, which
implies the $\tilde{S}^{(\gh )}$ factor must be restricted as before
using
the $\uqH$ invariance.
The $Q^{(k)}  , \qb^{(k)}$ symmetries are
unaffected. For $\be = \be^{(H)}_-$,
$1/ \gamma_k = 1/(l+h)$ is independent of $k$,
and $q$ is still given by \eIIxx .
This leads to the following result for the
perturbed coset theories.
The spectrum consists of kinks with a  RSOS $\ot$ RSOS structure
\eqn\eIIxxivb{
K^{(n)}_{\lambda^{(G)}_j \lambda^{(G)}_i  ;
\lambda^{(H)}_{j'} \lambda^{(H)}_{i'}
} (\th ) }
of mass $M_n$.  The S-matrix for these kinks is given by
\eqn\eIIxxv{
S^{(G/H);(k,l)}_{nm} (\th )  = X_{nm} (\th ) ~
\tilde{S} ^{(\gh /G);k}_{nm} (\th ) \ot
\tilde{S} ^{(\gh /H);l}_{nm} (\th ) . }

\newsec{Examples}

The simplest applications of our techniques is to the $\IC \IP^N$
models:
$$\CM_{N,k} ~\equiv ~ {{SU_k(N+1) \times SO(2N)} \over {SU_{k+1}(N)
\times U(1)}} \ .$$
The S-matrices were obtained in \rfii\ for the solitons in the
affine Toda perturbations of these models.  The results for the
$\CM_{N,1}$ models were derived by a considerable amount of hard
work  using the \LG structure and bootstrap methods.  The S-matrices
for the general $\CM_{N,k}$ models were then conjectured and a
compelling body of evidence was presented.  Here we will derive, in
a rather straightforward way, the
general S-matrix for the $\CM_{N,k}$ models.  Our techniques have
the
advantage that we make explicit use if the underlying $H$-Toda
structure, and we naturally incorporate the supersymmetry
through the extension to the affine $G$ Toda structure.
{}From the previous section we see immediately that the S-matrix of
the $\CM_{N,k}$ models with $k>1$ has the tensor product structure
that was advanced in \rfii.  We therefore only need to examine the
factor $\tilde{S} ^{(\gh /H);g-h}_{nm} (\th )$ of \eIIxxv\ in more
detail, or equivalently, analyze the scattering matrices for
the $\CM_{N,1}$ models.

For later convenience, consider the $\check R$-matrix for the
fundamental vector ($(M+N)$-dimensional) representation of $SU(M+N)$.
In the homogeneous gradation this is given by\rJimcmp
\eqn\checkrmat{\eqalign{\check R(x,q) \ = \ & (xq-x^{-1}q^{-1})
\sum_{\alpha=1}^{M+N} E_{\alpha \alpha} \otimes E_{\alpha \alpha}
+ (x- x^{-1}) \sum_{ { \alpha \neq\beta\atop \alpha,
\beta = 1 }}^{(M+N)} E_{\alpha \beta} \otimes E_{\beta \alpha} \cr
& +  (q-q^{-1} )\Bigl [ x \sum_{ \alpha > \beta}  ~+~ x^{-1}
\sum_{\alpha < \beta} \Bigr ] E_{\alpha \alpha }
\otimes E_{\beta\beta } \ , \cr }}
where  $E_{\alpha\beta}$ is an $(M+N)\times (M+N)$ matrix whose
entries
$(E_{\alpha \beta})_{ij}$ are equal to $\delta_{\alpha i }
\delta_{\beta j}$.
The spectral parameter $x$ is determined from \eIIxi; at the
$N=2$ supersymmetric point it is $x= \exp (\th /g )$, where $g=M+N$.
Let the indices $a,b$ and $i,j$ run from $1$ to
$M$ and from $M+1$ to $M+N$ respectively.  Let $\check R_{(1)}(x,q)$
and  $\check R_{(2)}(x,q)$ be the diagonal $M \times M$ and
$N \times N$ blocks in $\check R$.
Note that the sub-matrices $\check R_{(1)}$ and $\check R_{(2)}$
are simply the $\check R$-matrices for $SU(M)$ and $SU(N)$
respectively.  Now perform the conjugation operation by
$\sigma_G^{-1} \ot \sigma_G^{-1}$ to pass to the principal gradation,
and then perform the conjugation \eIIxv.  Let $\ckpR$ denote the
matrix that results from these two conjugations.  The combined effect
of the two conjugation operations does not modify the
structure of the sub-matrices
$\check R_{(1)}$ and $\check R_{(2)}$, except that
$x\to x^{g/2}$.  One also finds that the
off-diagonal blocks of $\ckpR$ become extremely simple:
\eqn\offdiag{\eqalign{\check R^\prime(x,q)_{a i, b j} &~=~
\check R^\prime(x,q)_{i a, j b} ~=~ (q ~-~ q^{-1}) ~\delta_{ab}
{}~\delta_{ij} \cr  \check R^\prime(x,q)_{i a, b j} &~=~
\check R^\prime(x,q)_{a i, j b} ~=~ (x^{g/2}  ~-~ x^{-g/2}) ~\delta_{ab}
{}~\delta_{ij} \ . \cr}}

It is elementary to convert this to the required RSOS matrix.  The
Fock space, $\CF$, in \eIIxvii\ is obtained by tensoring together
fundamental representations of $G$ and decomposing into highest
weights of $H$. These $H$ representations are then restricted to the
affine highest weights of $H_{g-h}$.
Let $\Lambda \equiv (\lambda, \nu; q)$ denote a weight of $SU(M+N)$
that decomposes into an affine highest weight $\lambda$ of $SU_N(M)$,
an affine highest weight $\nu$ of $SU_M(N)$ and let
$q = q_{U(1)}$ $ = \frac{2}{g} (\rho_G - \rho_H) \cdot \Lambda$ be
the
$N=2$, $U(1)$ charge of the vector $\Lambda$.  Now consider the
matrix elements $(\ckpR)^{\Lambda_1, \Lambda_2}_{\Lambda_4,
\Lambda_3}$
that are involved in the exchange relation \eIIxix.  Let
$\Lambda_i \equiv (\lambda_i, \nu_i; q_i)$.  For such a
matrix element to be non-zero there are three possibilities:
\hfill\break
\noindent
(i) The exchange does not involve the $SU(N)$ factor; {\it i.e.}
$\nu_1$ $=\nu_2$ $ =\nu_3$ $=\nu_4$ $=\nu$.
\hfill\break
\noindent
(ii) The exchange does not involve the $SU(M)$ factor; {\it i.e.}
$\lambda_1$ $=\lambda_2$ $ =\lambda_3$ $=\lambda_4$ $=\lambda$.
\hfill\break
\noindent
(iii) The off-diagonal terms of \offdiag\ act.
\smallskip
\noindent
The expression for $\ckpR$ in situations (i) and (ii) is well
known ( see, for example, \ref\FZU{P.~Di~Francesco and J.B.~Zuber,
\nup{338} (1990) 602.} ).
In situation (ii), one can write
\eqn\rsosR{\ckpR ~=~ \sinh \Big(\frac{i \pi}{M+N} ~-~ \frac{\th}{2}
\Big)
{}~+~ \sinh \Big( \frac{\th}{2} \Big) ~\CU \ ,}
where we have taken $ q = - e^{-i \pi/(M+N)}$ and used
$x^{g/2}  = e^{\th /2}$.  To define $\CU$, introduce the vectors
$e_1 \equiv \xi_1$, $e_N \equiv  - \xi_{N-1}$
and $e_j \equiv \xi_j - \xi_{j-1}$  for $ j=2, \ldots , N-1$,
where $\xi_j$ is the $j^{th}$ fundamental weight of  $SU(N)$,
and define a function
\eqn\defsjk{ s_{jk}(\nu) \ \equiv  \ \sin \Bigl ( {\pi \over M+N}
(e_j-e_k) \cdot (\nu + \rho) \Bigr )  \ . }
where $\rho$ is the Weyl vector of $SU(N)$.
The operator, $\CU$, then has the form:
\eqn\defup{ \CU \ \equiv \ (1-\delta_{jl}) {\Big( s_{jl}(\nu + e_j)
s_{jl}(\nu + e_k)\Bigr )^{\half} \over s_{jl}(\nu) } \ ,  }
where $\Lambda_{1} \equiv (\lambda, \nu ; q )$,
$\Lambda_2 \equiv (\lambda, \nu + e_j; q + \frac{M}{M+N})$,
$\Lambda_{3} \equiv (\lambda, \nu + e_j + e_l; q + \frac{2M}{M+N})$
and $\Lambda_4 \equiv (\lambda, \nu  + e_k;q + \frac{M}{M+N})$.
The RSOS reduction of the $\ckpR$ matrix in the $SU(M)$ factor is
much
the same.  Note that for $M=1$, the $\ckpR$ matrix in the ``$SU(1)$''
direction involves only the $U(1)$ charge and, as can be seen from
\checkrmat, the $\ckpR$ matrix reduces to a simple multiplicative
factor of $\sinh (\frac{\th}{2} - \frac{i \pi}{M+N})$ \ ({\it i.e.}
$\CU$ vanishes).

In the foregoing components of $\ckpR$ one either had $\lambda_1
= \lambda_2$ $= \lambda_3 $ or $\nu_1 = \nu_2$ $= \nu_3 $.  The other
non-zero components come from the off diagonal blocks, and one has
\eqn\redoffR{\ckpR ~=~ \cases {i\ \sin \big( \frac{\pi}{M+N} \big)
& or \cr \sinh \big( \frac{\th}{2} \big) &   \cr}}
where one has $i~\sin (\pi/M+N)$ for either a) $\Lambda_1 = (\lambda_1,
\nu_1; q)$, $\Lambda_2 = \Lambda_4  = (\lambda_1, \nu_2;
q + \frac{M}{M+N})$ and $\Lambda_3 = (\lambda_3, \nu_2; q +
\frac{M-N}{M+N})$, or b) $\Lambda_1 = (\lambda_1, \nu_1; q)$,
$\Lambda_2 = \Lambda_4  = (\lambda_2, \nu_1;
q - \frac{N}{M+N})$ and $\Lambda_3 = (\lambda_2, \nu_3; q +
\frac{M-N}{M+N})$; and one has $\sinh (\th/2)$ for either c)
$\Lambda_1 = (\lambda_1, \nu_1; q)$, $\Lambda_2 = (\lambda_1, \nu_2;
q + \frac{M}{M+N})$, $\Lambda_3 = (\lambda_3, \nu_2; q +
\frac{M-N}{M+N})$ and $\Lambda_4 = (\lambda_3,\nu_1;q -
\frac{N}{M+N})$,
or d) $\Lambda_1 = (\lambda_1, \nu_1; q)$,  $\Lambda_2  =
(\lambda_2, \nu_1;  q - \frac{N}{M+N})$, $\Lambda_3 = (\lambda_2,
\nu_3;
q +  \frac{M-N}{M+N})$ and $\Lambda_4 =
(\lambda_1,\nu_3;q - \frac{N}{M+N})$.

One should also note that the fundamental, $(M+N)$-dimensional
representation of $SU(M+N)$ decomposes into the
$(M,1)(-\frac{N}{M+N})$
$\oplus (1,N)(+\frac{M}{M+N})$ of $H = SU(M) \times SU(N) \times
U(1)$.
Moreover, these two representations of $H$ are mapped into one
another
by the generators $X_{\pm \gamma}$ and $X_{\pm \psi}$ that extend
$\uqH$
to $\uqGh$.  This means that kink operators corresponding to
$(M,1)(-\frac{N}{M+N})$ and $(1,N)(+\frac{M}{M+N})$ should be a
doublet of the superalgebra.

Consider now the $\IC \IP^N$ models ({\it i.e}  set $M=1$ ).
Let $u_1$ and $d_1$ denote the kinks corresponding to the
$N(\frac{1}{N+1})$ and  $1(-\frac{N}{N+1})$ of $SU(N) \times U(1)$.
As in \rfii\ these are a doublet of solitons of mass $M_1$.  Note
that $u_1$ has fermion number $\frac{1}{N+1}$ and $d_1$ has fermion
number $-\frac{N}{N+1}$.  The RSOS heights are restricted to
affine highest weights of $SU_1(N) \times U(1)$ and one can easily
see from \defsjk\ and \defup\ that $\CU$ is non-zero if and only if
$\nu = \xi_{j-1}$, $l = j+1$ (mod $N$) and $\Lambda_2 = \Lambda_4$.
One then has $\CU = 2 \cos (\pi/(N+1))$ and hence \rsosR\ gives a
factor of
$-\sinh (\th/2 - i\pi/(N+1)) + 2  cos(\pi/(N+1)) \sinh (\th/2) =
\sinh (\th/2 + i\pi/(M+1))$.  Consequently the scattering matrix is:
$$\eqalign{u_1 \ u_1  ~&\rightarrow~ u_1 \ u_1 \ \ \ \ \qquad\qquad
Z_{1,1} \sinh \Big(\frac{\th}{2} ~+~ \frac{i \pi}{N+1} \Big) \cr
d_1 \ d_1  ~&\rightarrow~ d_1 \ d_1 \ \ \ \ \qquad \qquad -Z_{1,1}
\sinh \Big(\frac{\th}{2} ~-~ \frac{i \pi}{N+1} \Big) \cr}$$
$$ \left.\eqalign{  u_1 \ d_1  ~ \rightarrow~ u_1 \ d_1 \cr d_1 \ u_1
{}~ \rightarrow~ d_1 \ u_1  } \right\} \qquad \qquad
i \ Z_{1,1}
\sin \Big(\frac{\pi}{N+1}  \Big) \qquad \qquad $$
$$ \left.\eqalign{  u_1 \ d_1  ~ \rightarrow~ d_1 \ u_1 \cr d_1 \ u_1
{}~ \rightarrow~ u_1 \ d_1  } \right\}
\qquad \qquad  Z_{1,1}  \sinh \Big(\frac{\th}{2}
\Big)\qquad \qquad \qquad $$
where $Z_{1,1} = X_{1,1} v_{1,1}(\th, q)$, and
$$X_{1,1} = \frac{ \sin \( \frac{\th}{2i} + \frac\pi{N+1} \) }
{ \sin \( \frac\th{2i} - \frac\pi{N+1} \) } $$
\
$$v_{1,1}
= \inv{2\pi i}
\Gamma \( \frac{i\th}{2\pi} + \inv{N+1} \)
\Gamma \( 1- \frac{i\th}{2\pi} - \inv{N+1} \)
\prod_{j=1}^\infty
\frac{ \Gamma \( 1+ \frac{i\th}{2\pi} + j -1 \) }
{\Gamma \( 1-\frac{i\th}{2\pi}
+ j-1 \) }$$
$$\times
\frac{ \Gamma \(  \frac{i\th}{2\pi} + j  \) }
{\Gamma \( -\frac{i\th}{2\pi}
+ j \) }
\frac{ \Gamma \(  -\frac{i\th}{2\pi} + j-1 +\inv{N+1}  \) }
{\Gamma \( \frac{i\th}{2\pi}
+ j-1 +\inv{N+1} \) }
\frac{ \Gamma \( 1 -\frac{i\th}{2\pi} + j- \inv{N+1}  \) }
{\Gamma \( 1+ \frac{i\th}{2\pi}
+ j - \inv{N+1} \) } .$$
 This S-matrix
 agrees with the results in \rfii.  The rest of the
scattering amplitudes can be deduced by conjugating and making RSOS
reductions of the $\check R$ matrices for other fundamental
representations of $G = SU(N+1)$.
Alternatively, they can be obtained by fusing the S-matrix given
above.

While one can directly relate the scattering matrices of the affine
$\hat G$ Toda theory to the scattering matrices of the \LG solitons,
it
is important to point out some of the apparent differences between
the
two theories and how these differences could possibly be resolved in
mapping one theory onto the other.  The most obvious difference is
that
the solitons of the \LG theory interpolate between the finite number
of \LG vacua, whereas the kinks of the affine Toda theory interpolate
between the infinite set of affine $H_{g-h}$ highest weights that
appear
in tensor products of $G$ representations.  In particular these
highest
weights are unrestricted in the range of charges in the free $U(1)$
factor of $H$.    The simplest example of this is the $\CM_{1,1}$
model with $c=1$.  The \LG potential is  $W(x) = {1 \over 3} x^3 -
a^2 x$,
with vacua at $x = \pm a$.  The associated affine Toda theory is
simply
sine-Gordon theory at  choice of the coupling constant that yields a
supersymmetric model.  The latter theory has infinitely many
distinct vacua.  However, if one considers the allowed
scattering between solitons in  both theories one comes up with
the same scattering matrix, and a  natural identification of \LG
solitons and sine-Gordon kinks.  One should
also remember that in the $c=1$, $N=2$ superconformal theory the
supersymmetry generators are:
$$ G^\pm (z) ~=~ e^{\pm i \sqrt{3} \phi(z)} \ ,$$
while the chiral primary field $x(z)$ is given by
$$  x(z) ~=~ e^{{i \over \sqrt{3}} \phi(z)} \ .$$
In the ultra-violet limit of the sine-Gordon theory the fundamental
kink operator reduces to:
\eqn\sgkink{K^\pm(z) ~=~ e^{\pm {i \sqrt{3} \over 2} \phi(z)} \ .}
The kink fields corresponding to \sgkink\ has $\Delta \phi
\equiv \phi(+\infty) - \phi(-\infty) = \pm \pi \sqrt {3} $.
Observe that if one shifts $\phi$ by this amount then one maps
$x(z) \rightarrow -x(z)$.  A shift of $\pm 2 \pi \sqrt{3}$ (which
corresponds to the supercharge since $K^\pm(z) K^\pm(w)
\sim (z-w)^{3/4} G^\pm(z)$) maps $x(z)$ back to itself.
Thus, for the choice of the coupling constant $\beta = \sqrt{2/3}$,
one can view the ``double-kink''operators ({\it i.e.} the kink
operators with $\Lambda = \pm \sqrt{2}$) as a supercharges.
Incorporating these supercharges into the chiral algebra
of the conformal model thus
has an off-critical counterpart in which one can replace the
sine-Gordon theory by a supersymmetric effective field theory of
the operator $x(z)$.  The sine-Gordon kinks then map onto
\LG solitons running between $-a$ and $+a$.  One can then use the
vertex
operator realization of the supercharges to see how the sine-Gordon
kinks fall into supermultiplets of \LG solitons.

The generalization of the foregoing to less trivial models is made
much more difficult by the necessity of screening.  One can try to
use the vertex operator realization of the supercurrents to map
the infinitely many vacua labelled by the highest weights of
affine $H_{g-h}$ to the finitely many vacua of the \LG theory.
One can then attempt to map the Toda kinks onto \LG solitons.
This can be carried out successfully in the $\CM_{N,1}$
models.  The basic difficulty is that the screenings and BRST
reduction make it  difficult to determine exactly how
certain Toda kink operators, or perhaps combinations of Toda
kink operators, interpolate between \LG vacua.  There are, however,
certain elementary reductions that can be made in general. For
example observe that shifting $\phi$ by $\frac{2 \pi}{\beta}
g \Lambda =  2 \pi \sqrt{g(g+1)}\Lambda$ for $\Lambda \in M^*(G)$
will not change the chiral primary fields.  The RSOS reduction of
the S-matrix will also not change under such a shift.  This means
that if one incorporates suitable $U(1)$ translations on the
weight lattice of $G$ into the chiral algebra then one can reduce
the kink spectrum of the Toda theory to a finite subset.  Then, if
one can properly incorporate the supercharge one should be able to
further reduce this finite subset to the \LG solitons.
We have already shown how to
extract the  correct \LG soliton
scattering matrix from the affine Toda scattering matrix in the
$\IC\IP^N$ models, and this suggests it is indeed possible
to make the desired
identifications of Toda and \LG solitons.

It is desirable however
 to find an {\it a~priori}
argument based on the supersymmetry that shows how to deduce
the \LG spectrum from the Toda spectrum.
The ensuing difficulties in doing this are best illustrated by  a
simple
example.  Consider the model
\eqn\sufour{{{SU_1(4) \times SO_1(8) } \over {SU_2(3) \times SU_3(2)
\times U(1)}}\ . }
This is the $c=12/5$ minimal model, but with the type $D$ modular
invariant.  The perturbed \LG superpotential is
$$W(x,y) ~=~ \frac{1}{5} x^5 ~+~ xy^2 ~-~ a^4 x \ . $$
The critical points are at $x=0$, $y =\pm a^2$ and at $y=0$,
$x=\pm a, \pm ai$.  A schematic representation of the soliton
polytope \rlg\ is shown in figure 1, and the mass projection
\rntwo\ is shown in figure 2.

\epsfxsize = 3.5in
\vbox{\vskip -.1in\hbox{\centerline{\epsffile{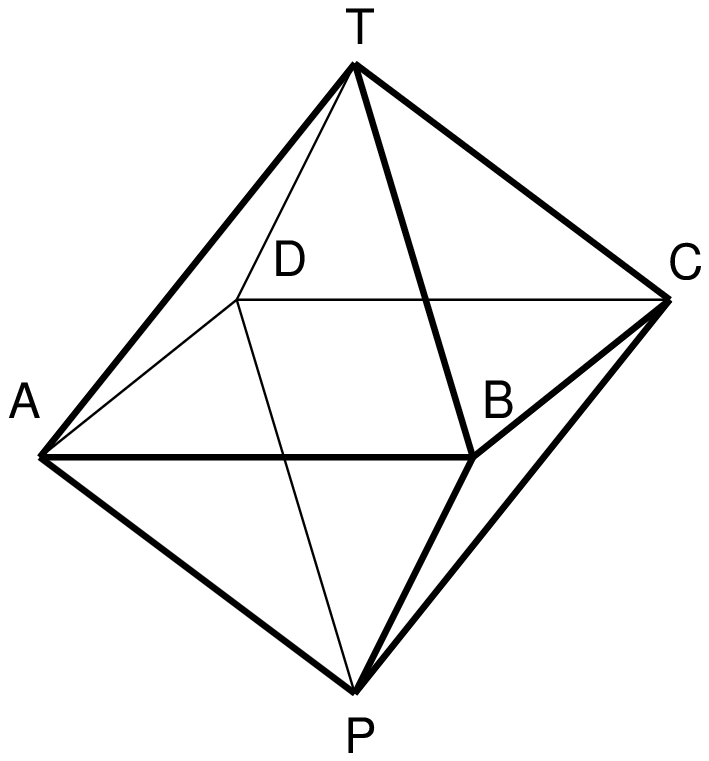}}}
\vskip -.5in
{\leftskip .5in \rightskip .5in \noindent \ninerm \baselineskip=10pt
Figure 1. The soliton polytope for the integrable model with
\LG potential $W(x,y) ~=~ \frac{1}{5} x^5 ~+~ xy^2 ~-~ a^4 x \ . $
\smallskip}} \bigskip \bigskip

The top (T) and bottom (P) vertices
of the soliton polytope project to the center of the square in
the mass projection.  Each  line from the center of the square
to a corner thus represents two distinct types of soliton
supermultiplet:
ones that start (or finish) at T and ones that start (or finish) at
P.
Let $M_1$ be the mass of these solitons and let $M_2$ be the mass
of the solitons that run along the edges (from corner to corner)
of the square in figure 1b.  From the results of \rntwo\ we
know that $M_2/M_1 = \sqrt{2}$.   Label the corners of the square by
A, B, C
and D.  The geometry suggests that scattering
the soliton that runs from A to T
against a soliton that runs from T to B should give a resonance at
$\th = i
\pi/2$ for creating, at rest, a soliton that runs from A to B.  These
facts
lead us to associate the solitons of mass $M_1$ with the $4$ or
$\bar 4$  of
$SU(4)$ and those of mass $M_2$ with the $6$ of $SU(4)$.  It was
noted in
\rntwo\ that the foregoing solitons could not form a closed
scattering theory
since scattering a soliton running from T to A against a soliton from
A to B
must yield an outgoing state of mass $M_2$ that starts at T.  There
are no
fundamental \LG solitons that satisfy this, but if one added particle
or
breather states of mass $M_2$ that are localized at P and T then one
could
probably close the scattering theory.

\epsfxsize = 3.0in
\vbox{\vskip -.1in\hbox{\centerline{\epsffile{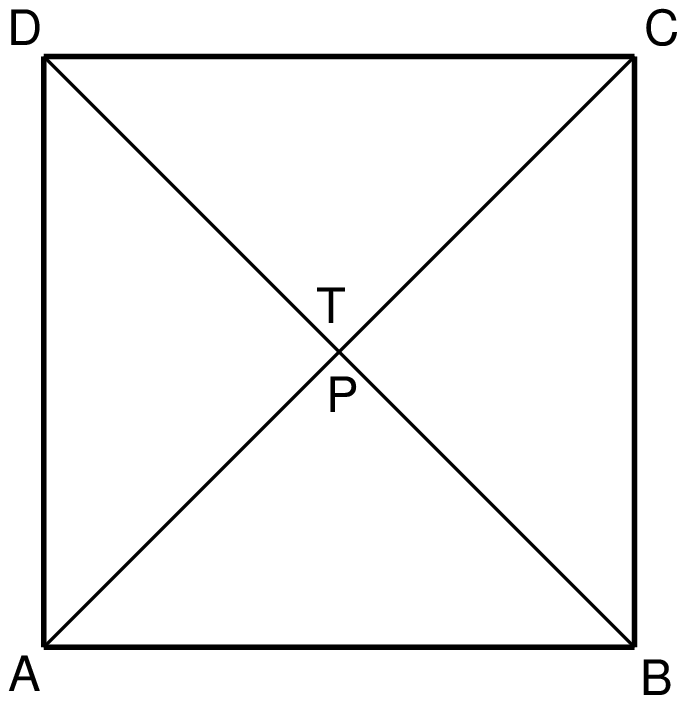}}}
\vskip -.3in
{\leftskip .5in \rightskip .5in \noindent \ninerm \baselineskip=10pt
Figure 2. The mass projection of the soliton polytope for the
\LG potential $W(x,y) ~=~ \frac{1}{5} x^5 ~+~ xy^2 ~-~ a^4 x \ $.
The lines from the center to the corners correspond to two types of
solitons: those that connect to T and those that connect to P.  Both
these types of soliton have mass $M_1$.  The solitons running between
the corners have mass $M_2 = \sqrt{2} M_1$.
\smallskip}} \bigskip \bigskip

Our analysis provides a candidate S-matrix for the foregoing model.
The
problem is to see how the affine Toda theory relates to the \LG
theory.  The
soliton masses certainly agree.  One can also verify that the kinks
have the
correct anomalous fermion numbers.  The $4$ of $SU(4)$ decomposes
into
$(\frac{1}{2},0)(-\frac{1}{2}) \oplus (0, \frac{1}{2})(+\frac{1}{2})$
of $SU(2)
\times SU(2) \times U(1)$, while the $6$ decomposes into $(0,0)(-1)
\oplus
(\frac{1}{2},\frac{1}{2})(0) \oplus (0,0)(+1)$.  One can check this
against the
formula given in \rfii\ref\jgfw{J.~Goldstone and F.~Wilczek, {\it
Phys. Rev.
Lett. }{\bf 47} (1981) 986; A.J.~Niemi and G.W.~Semenoff, {\it Phys.
Rep} {\bf 135}
(1986) 99.}:
\eqn\frfnum{f ~=~ \frac{1}{2 \pi} \Delta Im\bigg ( det \bigg(
{{\partial^2 W}
\over {\partial x_i \partial x_j}} \bigg) \bigg)  \quad {\rm mod} \ \
1 \ ,}
which yields the fermion numbers of $\frac{1}{2}$ mod $1$ for the $4$
and $\bar
4$ and $0$ mod $1$ for the $6$.  The peculiarities start to arise
when one
considers the supermultiplet structure.  The $6$ appears to decompose
into an
irreducible {\it three} dimensional supermultiplet, which is
impossible in the
standard representations of supersymmetry.  One might hope that this
problem
would be resolved by considering the representation of the
supersymmetry on the
full set of RSOS kinks \eIIxviii, but we find that the problem
persists.  One
tends to find that completely different kinks want to have a third
kink as a
common superpartner.  This  surprise can be elucidated somewhat by
taking the
ultra-violet limit of the perturbed model and looking at the limit of
the kink
operators in the conformal theory.  The affine Toda kink operators
limit to
operators of the form
$$ exp(- \frac{i}{\beta} \Lambda \cdot \phi(z) )$$
for $\Lambda$ a fundamental weight of $SU(4)$.  The original
superconformal
model restricted the labels $\Lambda $ to be of the form $w(\rho_G) -
\rho_G$,
which is always at least a root of $G=SU(4)$.   This restriction
amounted to
choosing a special, and indeed exceptional, modular invariant for the
numerator
factor of $H= SU(2) \times SU(2) \times U(1)$.  The choice of this
exceptional
modular invariant made a very important change to the fusion rules.
For a
general choice of modular invariant, the supercurrents, which
correspond to the
operators $\Phi^{0, -\gamma}_0$ and $\Phi^{0,\psi}_0$, would have
non-trivial
braidings with other operators.  For the exceptional modular
invariant, the
operators corresponding to the supercurrents become simple currents
and can
therefore be incorporated into an extended chiral algebra for the
theory, {\it
i.e.} they become supercurrents in the standard sense of the word.
The kink
operators in the ultra-violet limit correspond to conformal fields
that are
excluded from  the special modular invariant, and indeed have
non-abelian
braiding relationships with the operators corresponding to the
supercurrents.
Thus the new feature that we encounter is that the affine Toda kinks
have
non-abelian
braiding relationships with the supercurrents and so generate a
highly
non-standard supermultiplet structure.  The only class of models
where this
does not happen is, in fact, in
the $\IC \IP^N$ models, where the supercurrents are still
simple currents even when the kink operators are included.  However,
even in
this situation the kinks have anomalous fermion numbers which lead to
non-trivial, though abelian, braidings of the kinks with the
supercharges.

The foregoing discussion raises two possibilities:  either there is
non-abelian braiding of the supercharges with the \LG solitons, or
the
non-abelian braiding of the supercharges is purely an artifact of the
Toda
description and is not present in the \LG formulation.  One possible
method for
getting around the non-abelian braiding is to attempt to identify
several
distinct Toda kinks with a single \LG soliton so that the
supercharges are once
again well-behaved members of an extended chiral algebra.  In the
example
\sufour\ we could find no such identifications that were also
consistent with
the scattering matrix.  As regards being an artifact of our
formulation, we
note that anomalous abelian braiding relationships are already
present in the
\LG formulation of the $\IC \IP^N$ models \rfii\ as a consequence of
the
anomalous fermion number of the kinks.  Moreover, if one considers
the \LG
formulation of the model \sufour, one can see from \frfnum\ that the
solitons
of mass $M_1$ have fermion number $\frac{1}{2}$.
If we also make the plausible
assumption that the kink field should be local with respect to the
perturbing
operators, one can then argue that in the ultra-violet limit the
resulting
operator in the conformal field theory will have non-abelian
braidings with the
supercharges.   We therefore suspect that once one goes
beyond the $\IC \IP^N$ models, the kinks of the $N=2$ supersymmetric
$\gh$-affine Toda models will exhibit  non-abelian braiding
with
the supercurrents.  If this is the case, we expect that
one should also encounter this
phenomenon in the \LG formulation.

\newsec{Conclusions}

There are two major roles for $\check R$-matrices in two-dimensional
models:
either they appear in scattering matrices or they appear in Boltzmann
weights
of exactly solvable models.  Perturbed conformal theories give a
method of
interrelating both of these applications.  Given an exactly solvable
model on
can take its continuum limit, at criticality, and obtain a conformal
field
theory.  For such conformal models there is always a relevant
perturbation that
leads to a  quantum integrable model.  Another $\check R$-matrix  can
then
be obtained from the scattering theory of this integrable model.  For
the
non-supersymmetric minimal series,
 the
energy perturbation of the $p^{th}$ minimal model yields an
integrable model
whose scattering matrix involves an RSOS  $\check R$-matrix\rrsg,
and this $\check R$-matrix
defines the  Boltzmann weights of a lattice model whose continuum limit
is the
$(p-1)^{th}$ minimal
model\ref\rpasii{V. Pasquier, Nucl. Phys. {\bf B295} (1988) 491.}.
One can also see this sort of
progression in the
$N=2$ superconformal models, their lattice counterparts and the
related quantum
integrable models.  In \ref\rnwii{Z. Maassarani, D. Nemeschansky, and
N. P. Warner, USC preprint USC-92/007.},   the lattice analogues of the $N=2$
superconformal
models were described, and exactly the same $\check R$-matrices were
used there
as have been used employed in this paper, except that in \rnwii\ one
took $q =
e^{i \pi/(g+1)}$ in order to get the RSOS lattice model Boltzmann
weights.

One  way to understand the distinction between the
$\check
R$-matrices of the lattice model and the soliton S-matrix is that,
for the
former, one is working with the quantum group structure associated
with the
denominator of the coset, whereas for the latter, one is working with
the
quantum group structure of the numerator factors, as was observed in
\rBLii.  This means that
for the
quantum integrable perturbations of the $G \times H/H$ models, the
extension of
$\uqH$ to $\uqGh$ involves requiring that the S-matrix commutes with
the
supersymmetry charges \susys, whereas in the lattice model the same
extension
means  that the Boltzmann weights commute with the perturbation
operators
\perts.

\bigskip

\centerline{Acknowlegements}

This work is supported in part by the National Science Foundation.
A.L. and N.W. are supported in part by the Alfred P. Sloan Foundation.
This work was started at the  Aspen Center of Physics.

\listrefs
\bye